\documentclass[fleqn,10pt]{wlscirep}
\usepackage{graphicx}
\usepackage{adjustbox}
\usepackage[utf8]{inputenc}
\usepackage[T1]{fontenc}
\usepackage[english]{babel}
\usepackage{rotating}
\usepackage{subfigure}
\usepackage{comment}

\usepackage[finalnew]{trackchanges}

\title{Human Gist Processing Augments Deep Learning Breast Cancer Risk Assessment}

\author[1]{Skylar W. Wurster}
\author[2, *]{Arkadiusz Sitek}
\author[1, *]{Jian Chen}
\author[3]{Karla Evans}
\author[4]{Gaeun Kim}
\author[5]{Jeremy M. Wolfe}
\affil[1]{The Ohio State University, Department of Computer Science and Engineering, Columbus, OH 43210, USA}
\affil[2]{IBM Watson Health, Cambridge, MA 02142, USA}
\affil[3]{University of York, Department of Psychology, York, YO 10 5DD, UK}
\affil[4]{Stanford University, Department of Bioengineering, Stanford, CA 94305, USA}
\affil[5]{Harvard University, Brigham \& Women's Hospital, Cambridge, MA 02142, USA}
\affil[*]{Corresponding authors: arek@ibm.com and chen.8028@osu.edu}

\usepackage{multicol}
\begin{abstract}
\textbf{Radiologists can classify a mammogram as normal or abnormal at better than chance levels after less than a second's exposure to the images. In this work, we combine these radiologists' gist inputs into pre-trained machine learning models to validate that integrating gist with a CNN model can achieve an AUC (area under the curve) statistically significantly higher than either the gist perception of radiologists or the model without gist input.}
\end{abstract}
\begin{document}
\begin{multicols}{2}

\flushbottom
\maketitle
%
%


\thispagestyle{empty}






{Recent developments in machine learning models, such as convolutional neural networks (CNNs), utilize large input datasets to allow automatic solutions for screening the life-threatening breast cancers at an early more curable stage}~\cite{yala2019deep}. However,
these automatic solutions are reliant on the use of large annotated datasets from clinical diagnosis as input~\cite{sitek2019assessing}. 
Gist processing, a human visual ability~\cite{oliva2006building}, has not been utilized in neural networks for screening mammograms
even though it allows the human visual system to rapidly extract meaningful statistical information~\cite{evans2013gist, evans2016half}.
Evans et al. have shown that radiologist experts can classify mammograms as normal or abnormal at above chance levels after less than one second’s exposure to the image~\cite{evans2013gist}. Radiologists can even detect the gist of abnormality with only 500 milliseconds of exposure to completely cancer-free mammograms of women who would not develop cancer for another three years~\cite{evans2016half, evans2019MammoPrior, evans2018gist}.

In this work, we hypothesize that a CNN (\textit{aka, machine intelligence}) plus the information from  radiologist experts (\textit{aka, human perceptual intelligence}) will provide more accurate results than either the radiologists or the CNN alone. Our goal is not to generate a novel CNN architecture for screening, but instead add radiologist response in input to models already used.

\begin{figure*}[t!hp]
\centering
\includegraphics[width=\linewidth]{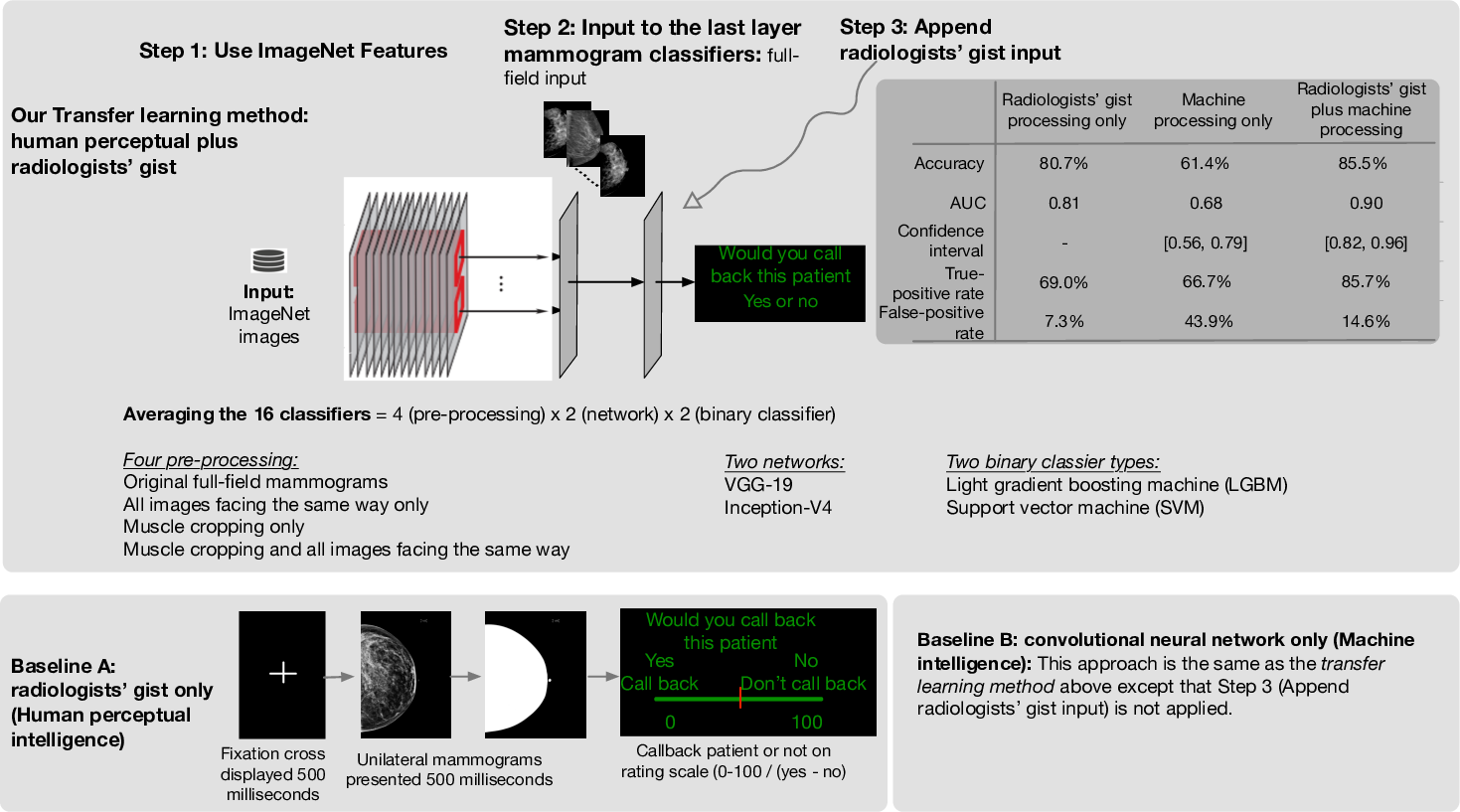}
\caption{Three testing methods used in our study and the resulting accuracy.}
\label{fig:method}
\end{figure*}

We use a transfer learning approach to combine radiologist gist information with features from a CNN classifier to investigate the benefits of coupling the machine and human expertise (Figure~\ref{fig:method}). 
Our method begins with the preprocessing of input dataset of 
mammogram images with four methods: (1) no changes,  (2) horizontally flip all left breast images, (3) crop muscle fibers out of each image, (4) combine (2) and (3), i.e., crop muscle fibers out and then horizontally flip all left breast images. When we horizontally flip left breast images, our resulting dataset will have images that face the same direction. 
Next, we feed each dataset into established CNNs 
Inception-v4 and VGG-19, pretrained on ImageNet, a corpus of over 14 million non-medical images~\cite{deng2009imagenet}. Though these networks are meant to classify non-medical objects.
we 
use deep features to provide an abstract representation from within the network to classify mammograms. 
Usually these features are unintelligible to humans, but some describe intuitive aspects of an image, such as edges, spirals, or gradients. In each network, we take all values from neurons of one layer near the final classification layer to obtain our feature vector.
Finally, we perform classification based on the feature vector obtained from the CNN using LightGBM (LGBM)~\cite{LGBM} and a linear support vector machine (SVM). 
Putting together, we have 4 processing steps of muscle cropping and rotation, two CNNs for deep features, and two classifiers for final classification, giving us $4*2 *2=16$ end-to-end systems. 

We test these 16 systems with and without appending the radiologist gist responses to the end of the feature vector before classification.
Original radiologists scores for detecting breast cancer were in range [0,100]. Radiologists' input is set to 0 (normal) if the average response is greater than 50 and to 1 (malignant) otherwise. We also calculate a ``confidence score'' from the radiologist gist input for each image:
$confidence = {|score-50|}/{50}$.

\textbf{Results}

\begin{figure*}[t!hp]
\centering
\includegraphics[width=\linewidth]{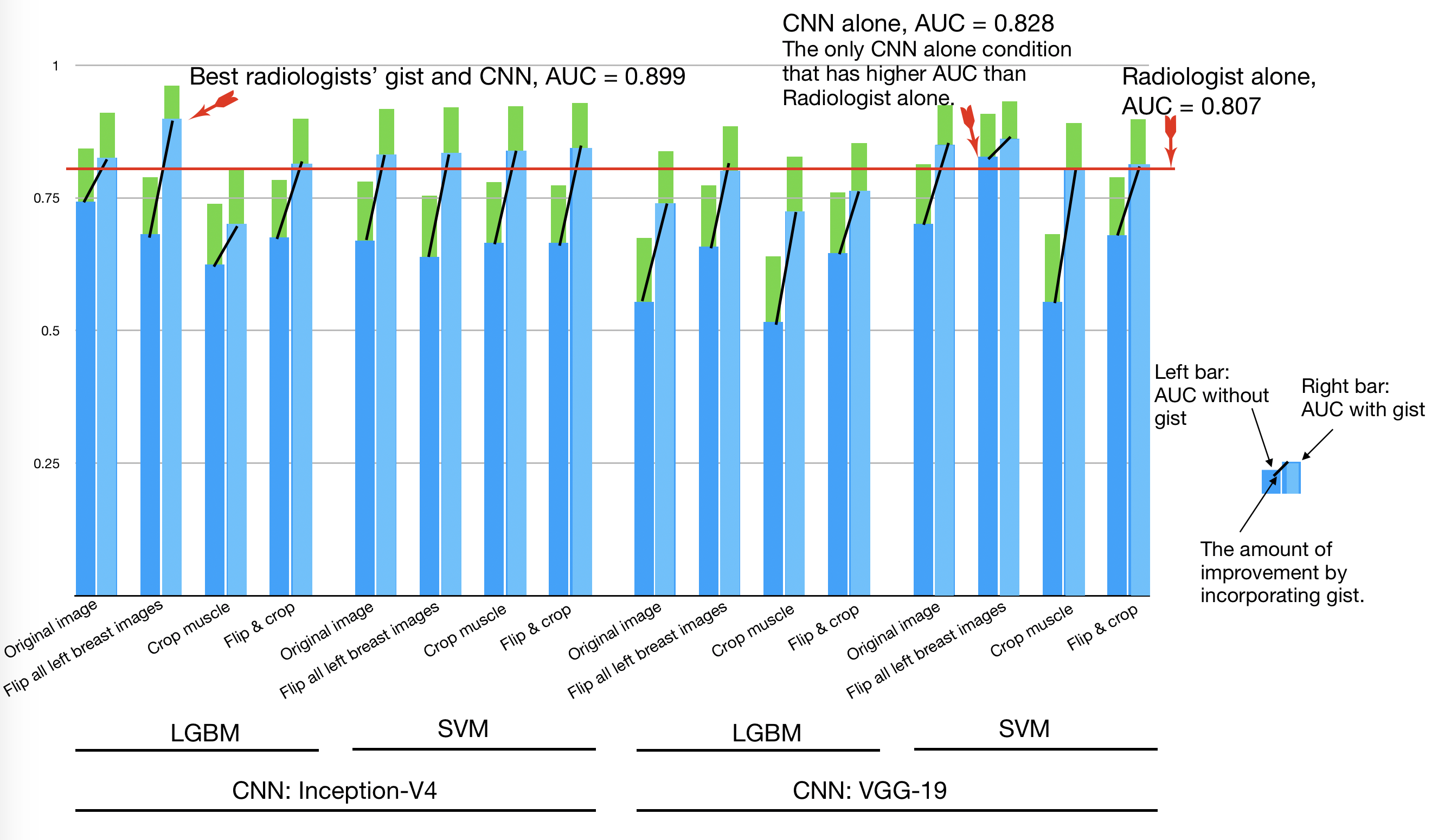} \\
\caption{AUC comparisons between the CNNs and radiologist gist plus CNNs.}
\label{fig:aucComparisons}
\end{figure*}

Our baseline condition (A) is the radiologists-alone condition in Evans et al.~\cite{evans2016half} (Figure~\ref{fig:method} lower-left): radiologist gist data are gathered by showing radiologists a unilateral mammogram of a mammogram image with no abnormalities, an image with an abnormality, or an image contralateral to an image with an abnormality for 500 milliseconds each. 
They perform at an above-chance level at classifying abnormal and normal mammograms in our dataset, with an accuracy of $80.7\%$ and an area under the receiver operating characteristic curve (ROC and AUC) of 0.81 (Figure~\ref{fig:aucComparisons}).

Our baseline condition (B) is the CNN-alone condition, when classifiers do not use radiologist gist response in the input vector (Figure~\ref{fig:method} lower-right and AUCs in Figure~\ref{fig:aucComparisons}, darker blue bars). Our end-to-end models on average achieve an AUC of 0.656.
Also, only one of the resulting AUCs (AUC = 0.828) from these 16 CNN-alone classifiers is better than the radiologist baseline of 0.809.
The $95\%$ confidence interval (CI) is $[0.739, 0.908]$) with the following system: preprocessing option 2 on inputs (input images facing the same direction), fed through VGG-19 for the features, and finally classified with a linear SVM. We think
overall this CNN-is-largely-worse-than-human may be due to our dataset being relatively small compared to other mammogram screening datasets.

Our new systems are to combine the baseline condition (B) with the radiologists' input (Figure~\ref{fig:method} top and AUCs in Figure~\ref{fig:aucComparisons} lighter blue bars). 
We see substantial improvements to the baseline B models when we add radiologist gist response into the classifier’s input in CNNs, with these new classifiers outperforming radiologists in some cases. 
The average AUC of the 16 classifiers increases to 0.814.
On average, each classifier improves its AUC by
0.157, reduces the false positive rate (FPR) by 0.188, improves the true positive rate (TPR) by 0.101, and improves accuracy by 0.144. In all cases, AUC, accuracy, true-positive rate (TPR), and false-positive rate (FPR) improve when adding radiologist gist response to the input vector, except for one that had a small decrease in TPR. In 10 of the 16 classifiers that use radiologist gist response in the input vector, we observe an AUC higher than the radiologists'. 
One result reports an AUC of 0.899, achieved with preprocessing option 2 (flip all left breast images) before feeding the inputs to Inception-v4 for the feature vector and then using LGBM for classification. 
Further, this stand-out case from our tests approaches the AUC of specifically trained deep network for the same task~\cite{IBMscreening}.

We further analyze this most successful classifier’s results before and after adding the radiologist gist response to the input vector.
We notice interesting trends when comparing the radiologists’ decisions to the classifier’s decisions. 
The classifier utilizing both the deep features and the radiologist gist input manages to correct 8 of the 16 false negative errors that the radiologists made, while only introducing three false positives and one false negative. On the other hand, after introducing the radiologist gist input to the classifier, 25 errors made by the classifier without gist response input are corrected, with only three errors introduced. These corrections include 8 false negatives and 17 false positives. There are also 5 instances in which both the classifier and the radiologists make the same mistake (see Supplementary materials for these image samples). 
In this same case, we see an improvement over the classifier not utilizing radiologist gist response, increasing AUC from
0.681($95\%$ CI [0.558, 0.789]) to 0.899($95\%$ CI [0.823, 0.962]). 
One-way Welch's \textit{F} test shows a statistically significant main effect of the classifier on the AUC 
($F(2, 1930)$ = $5117$, $p<0.0001$). A Games-Howell post-hoc test revealed that the AUC of the classifier without gist input is statistically significantly lower than the radiologists’ gist response (0.679 +/- 0.0597, $p<0.0001$), and that the classifier with gist input was statistically significantly higher than the radiologists’ gist response($0.897 +/- 0.0396$, $p<0.0001$).

Before introducing radiologist gist input into the classifier, we see a Pearson correlation coefficient of 0.113 between the classifier’s predictions and those of the radiologists. 
After introducing radiologist gist input into the classifier, we see a Pearson correlation coefficient of 0.733. Though these are highly correlated, the significant difference in AUC shows that the deep features from Inception-v4 or VGG-19 must capture additional information about the mammogram that radiologist gist responses do not, helping the classifier screen the mammograms with a higher AUC and accuracy than radiologist gist input alone. This is further supported by the fact that 10 of the 16 models using gist input appended to the feature vectors have higher AUC than radiologists’ gist response.

For the 16 radiologist errors, we see an average confidence of 0.178, lower than the overall average radiologist confidence of 0.346 ($p<0.05$).  Of the 8 mammograms that were incorrectly classified by the radiologist gist response and then corrected by the classifier, we observe an average confidence of 0.195, which is also lower than the average radiologist confidence ($p<0.05$).  Yet there are 9 predictions with radiologist confidence less than 0.2 in which the classifier is corrected when the radiologist gist input is present. This shows that the classifier is doing more than simply using high-confidence results provided by the radiologist gist response and using the deep features to classify low-confidence results. Instead, a more complex relationship has been learned between the deep features and radiologists’ gist input.

\textbf{Conclusion}

Our approach incorporates the informed decisions of radiologists’ who have years of education and experience with the image-analysis and pattern-recognition capabilities provided by deep CNNs and machine learning. The combination of the parts is better than either solution alone in our training data, as each input captures different features. Though helpful here, we suggest this combination may be used only for screening and not diagnosis since this approach cannot explain why it thinks there is apparent cancer. Similar solutions utilizing both inputs may prove useful in other problem domains, especially in the medical field where trained professionals often work with computer-aided detection systems and may pick up on different signals than does a deep neural network. In these cases, we suggest human and CNN collaborative problem solving.


\section{Supplementary Material}

\textbf{Methods}

\textbf{The mammogram input data.}
The mammography dataset contains 220 unilateral 
full-field digital mammograms obtained from 110 unique patients at Brigham and Women’s Hospital. These images fall into three classes: no malignancy (110 images), malignancy (66 images), or contralateral to the breast with a malignancy (44 images). All code and the dataset for this paper are on Github: https://gistmammocnn.github.io/.

\textbf{Radiologists' gist confidence scores.}
The radiologist data comprises responses from 10 radiologists who were each shown 120 images of the 220 where 40 were completely normal, 40 had a confirmed malignancy, and 40 were normal but contralateral to a breast with a malignancy. Observers were shown the image for 500 milliseconds and asked to report on a scale from 0, recommending the patient return for further examination, to 100, the scan is normal. This is experiment 2 from Evans et al.~\cite{evans2016half}. The readings from the
10 radiologists were averaged for each image to give the final radiologist response.

We compute a ``confidence score'' from each radiologist's response. Since the initial response is between 0 and 100, an average score of 50 means the radiologists were not confident one way or another. However, an average score near 0 or 100 means the radiologists were very confident there was either a malignancy or no malignancy. Therefore, we calculate the confidence score as the average response’s difference from 50, then normalized, and use this number as input in addition to the radiologist classification of 0 or 1. Then, scores such as 100 or 0 will have high confidence scores, and average scores near 50,  will have low confidence scores. Values of confidence in-between are interpolated: $confidence = {|score-50|}/{50}$.

\textbf{Data Preprocessing: Muscle Cropping and Image Orientation.}
It has been shown that cropping out pectoral muscle during pre-processing may be beneficial for classifiers when screening full-field mammograms for breast cancer. Due to our small dataset, we test four pre-processing techniques that involve
mirroring and cropping: (1) No crop – do not crop the pectoral muscle, and use the original full-field mammograms; (2) No crop, same direction – do not crop the pectoral muscle, but mirror all left breast mammograms, such that all mammograms are facing the same side. We expect this to reduce the abstraction that the classifier needs to learn during training, which it might not achieve with a small sample dataset; (3) Crop – crop pectoral muscle from the original mammograms by setting those pixel values to black; (4) Crop, same direction – crop the pectoral muscle from the original mammograms, and then mirror all left-breast images, for the same reasons as (2). We implement the cropping algorithm introduced in Rampun et al.~\cite{rampun2017fully} and visually inspect results to correct over- and under-cropping. 

\begin{figure*}[t!hp]
\centering
\includegraphics[width=\linewidth]{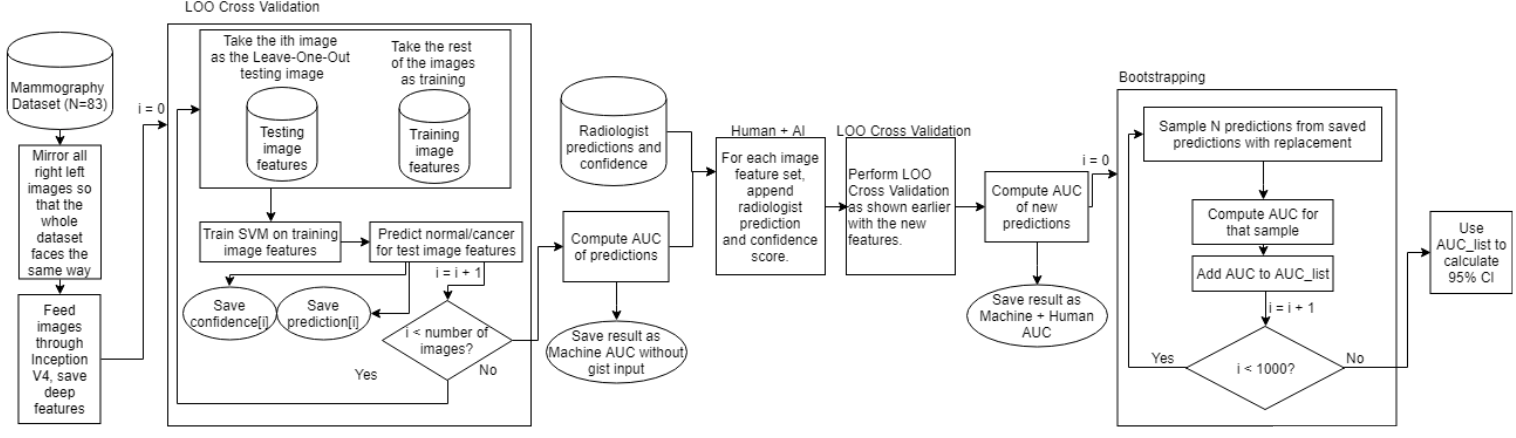}
\caption{Flow Diagram of Our Method.}
\label{fig:methodDetails}
\end{figure*}

\textbf{Transfer learning method.} Figure~\ref{fig:methodDetails} illustrates the transfer learning method we use. 
To generate deep features for the input images, we test two deep CNNs, VGG19 and Inception-v4. Both are pre-trained to classify over 1000 classes from the ImageNet corpus containing over 14 million non-medical images.  We use the python package pretrained models to load both models.
We take the values from nodes located in the first fully connected layer after the convolution layers to capture features about a full-field image fed through the network from Inception-v4, and use the values from the ReLu6 layer in VGG-19.  Thus, we have 1536 features from Inception-v4 and 4096 features from VGG-19. This creates a feature vector which is used as input into a supervised learning algorithm.  We test Microsoft’s LightGBM and a linear support vector machine (SVM) for binary classification as normal (0) or abnormal (1). LightGBM (LGBM) is a gradient-boosting framework using tree-based learning algorithms. We use the following parameters: learning rate = 0.003, boosting type = gbdt, objective = binary, metric = binarylogloss, subfeature = 0.5, numleaves = 10, mindata = 1, maxdepth = 1000. For the linear SVM, C=1000. Code is implemented in Python 3.6 with Keras and Tensorflow.

From the dataset, we use only the normal and malignant images for training and testing. Of these images, only those with radiologist responses are kept, leaving 83 images. This includes 42 abnormal mammograms and 41 normal. After the pre-processing step above, images were downscaled to
$299\times299$ pixels to fit the pre-trained Inception-v4 model properly via a function in pretrained models, or $224\times224$ to fit the pre-trained VGG19 model. Image pixel values are normalized as they are when Inception-v4 and VGG-19 are trained.
Leave-one-out (LOO) training and testing is used to generate 83 out-of-fold (OOF) predictions for each image in the set. Predictions will be 0 if classified as normal and 1 if classified as malignant.  With our four pre-processing methods, two deep CNNs, and two classifiers, we have 16 models for classification that do not use radiologist gist response input.

The above process is repeated to create another 16 models that incorporate the radiologist gist input by appending the radiologist gist classification (0 or 1) and the calculated \textit{radiologist gist confidence} (0.0-1.0), each multiplied by 100. This creates a new feature vector of length 1538 for each image when testing Inception-v4 feature vectors, and a feature vector length of 4098 when testing VGG19 feature vectors. We bootstrap 1000 AUC samples on the final predictions for each model to estimate the variance of the area under the receiver operating characteristic (AUC) for that model. A $95\%$ confidence interval is constructed using the middle
$95\%$ of samples from the bootstrapping set to account for any skew in the samples. Cohen’s \textit{d} is calculated using the mean difference divided by the standard deviation of the model. Significance between radiologist confidence subsets is tested using two-sample t-tests.

\begin{figure*}[t!hp]
\centering
\includegraphics[width=\linewidth]{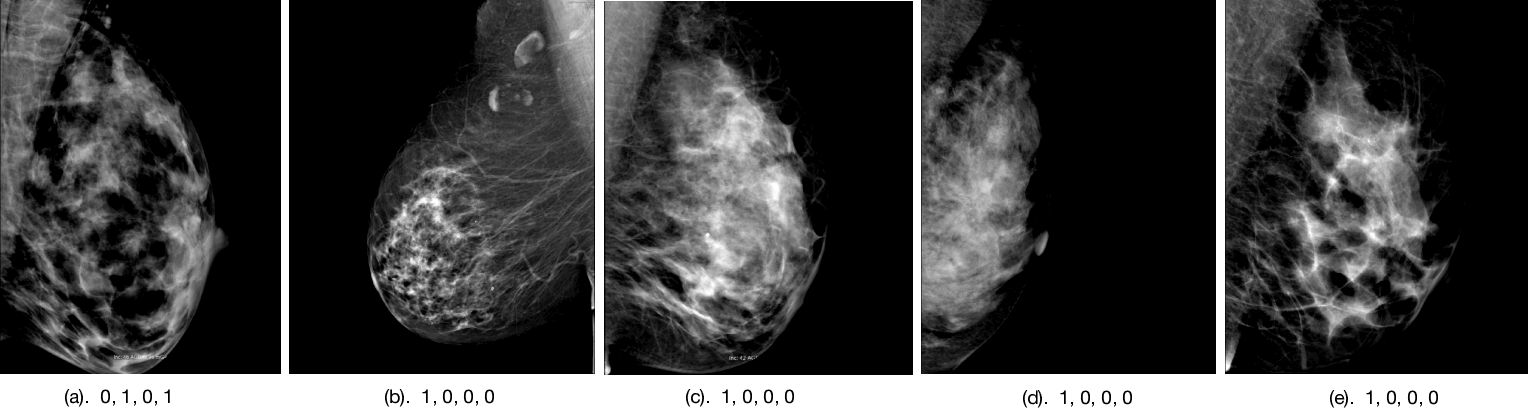}
\caption{Both humans and model make mistakes. The numbers under each images shows the \textit{ground truth}, \textit{radiologists' gist processing answer}, \textit{model without gist}, \textit{model with gist}. Here we select the numbers when item 2 and item 4 are not the same as item 1 (ground truth).}
\label{fig:BothWrong}
\end{figure*}

\begin{figure*}[t!hp]
\centering
\includegraphics[width=\linewidth]{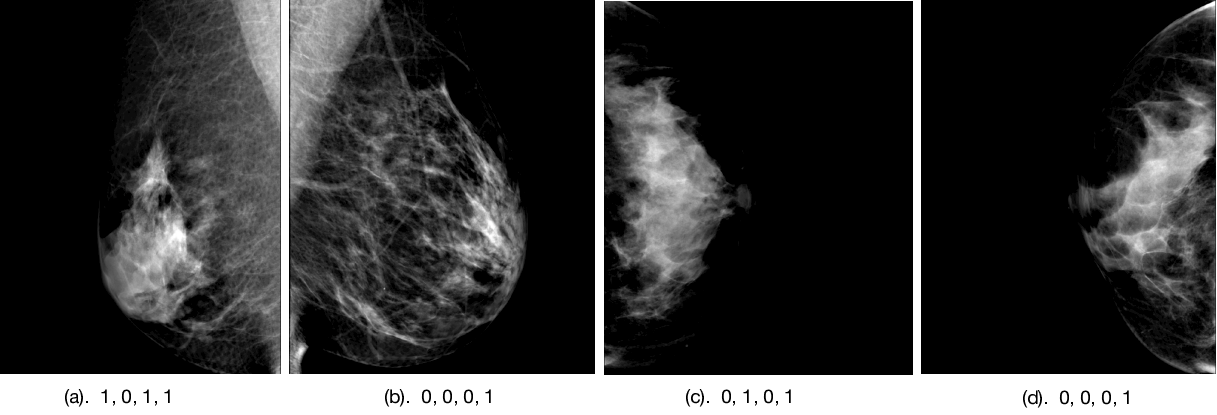}
\caption{Transfer learning model introduces mistakes. The numbers under each images shows the \textit{ground truth}, \textit{radiologists' gist processing answer}, \textit{model without gist}, \textit{model with gist}. In this case, item 1 and item 4 are different.}
\label{fig:modelIntroducesError}
\end{figure*}

\begin{figure*}[t!hp]
\centering
\includegraphics[width=\linewidth]{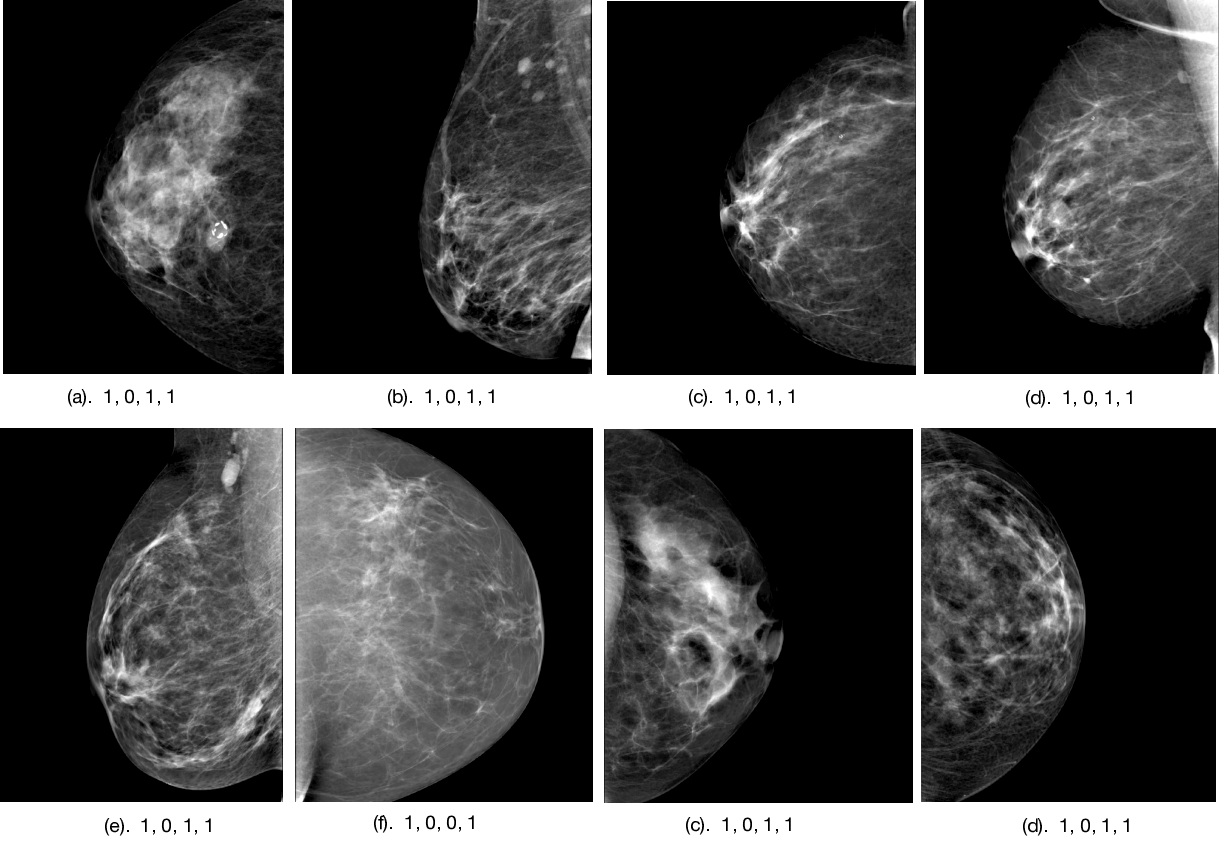}
\caption{Transfer learning model corrects radiologists' errors. The numbers under each images shows the \textit{ground truth}, \textit{radiologists' gist processing answer}, \textit{model without gist}, \textit{model with gist}. In this case, item 1 and item 2  are different and items 1 and 4 are the same.}
\label{fig:modelCorrectsError}
\end{figure*}

\begin{figure*}[t!hp]
\centering
\subfigure[]{\includegraphics[width=\linewidth]{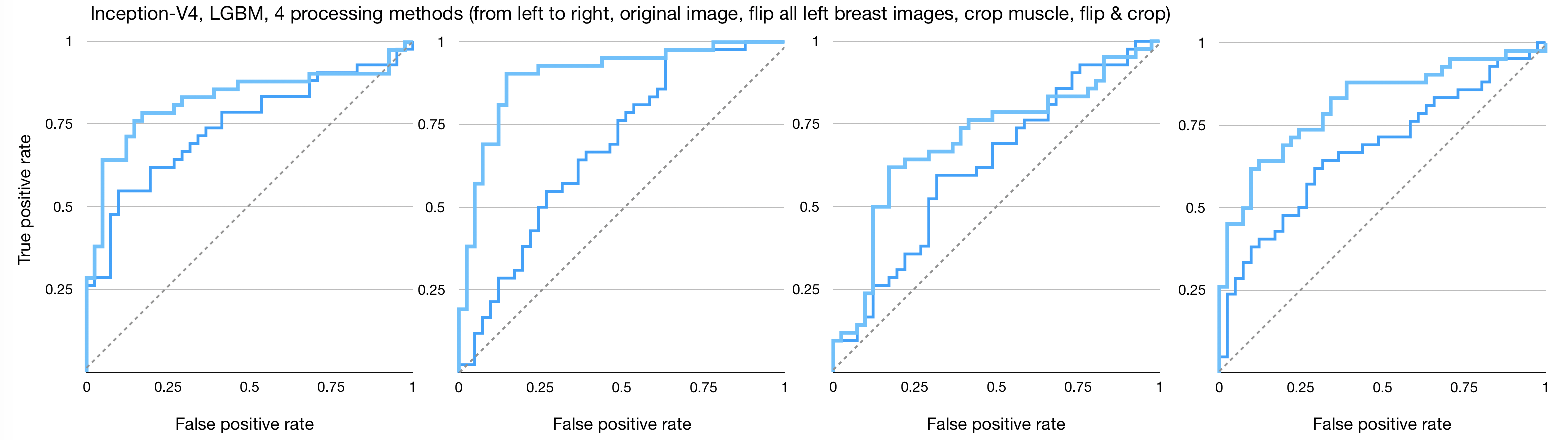}} \\
\subfigure[]{\includegraphics[width=\linewidth]{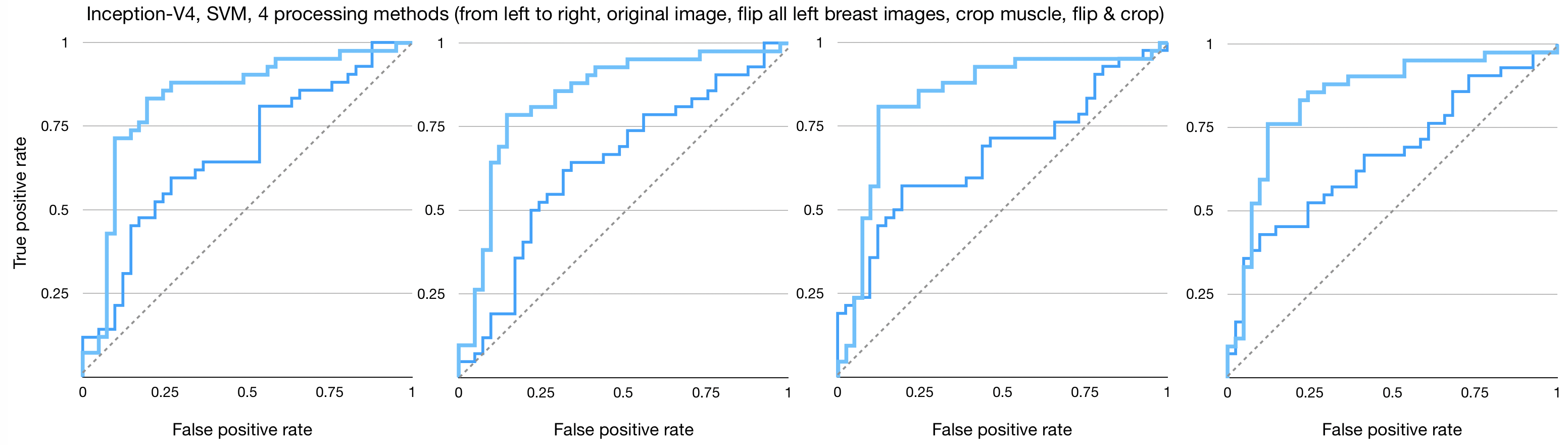}} \\
\subfigure[]{\includegraphics[width=\linewidth]{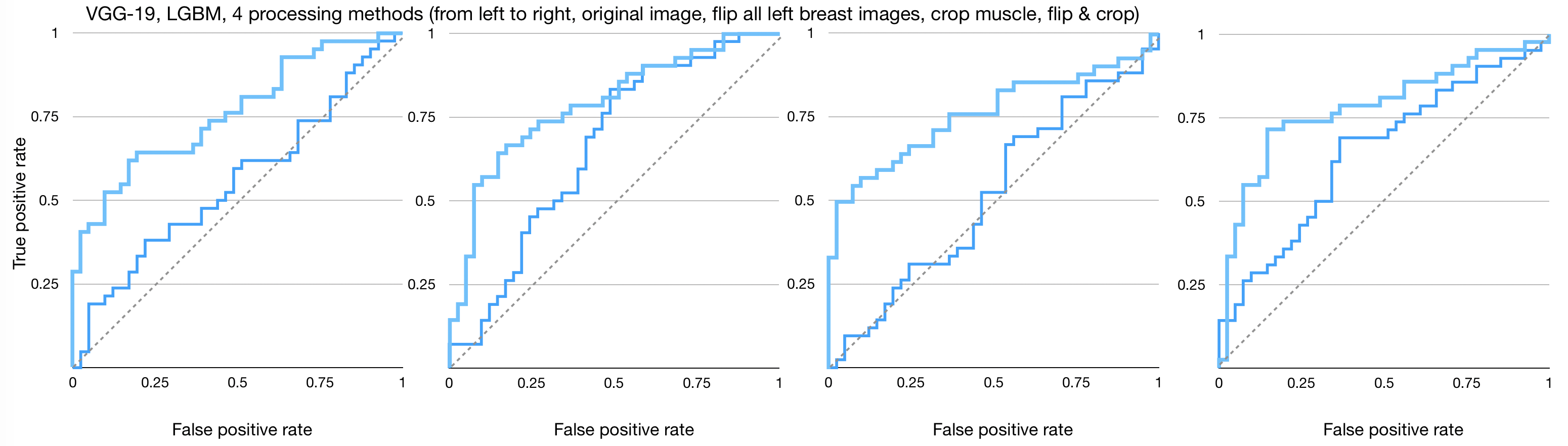}} \\
\subfigure[]{\includegraphics[width=\linewidth]{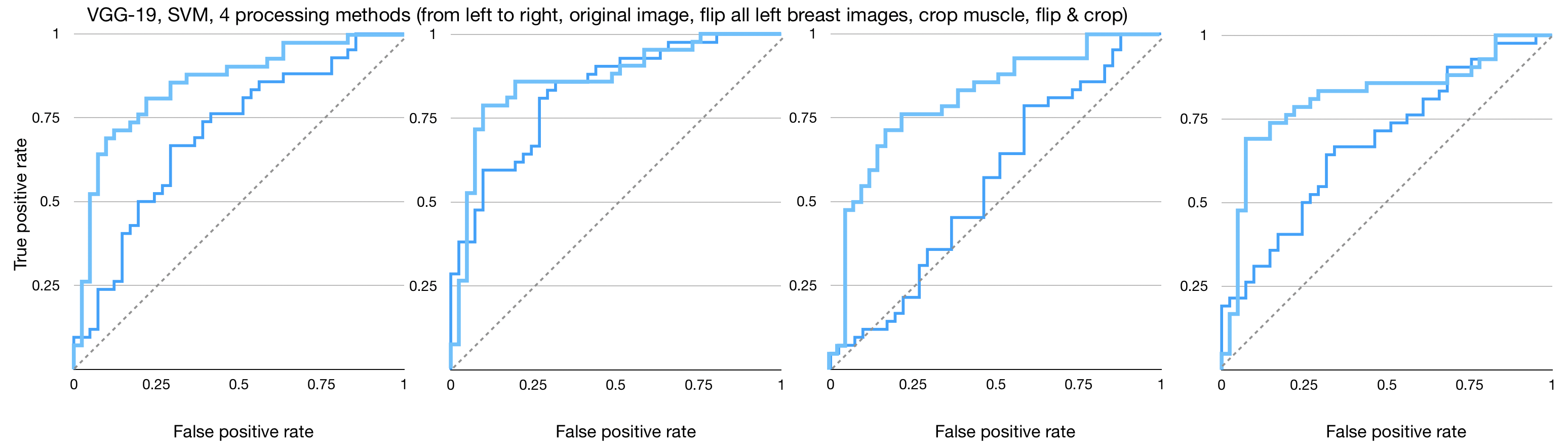}} \\
\caption{Receiver operating characteristic curve (ROC curve) for measuring Area Under the Curve (AUC) for each of the 16 end-to-end models: the CNNs (darker blue or lower AUC) and radiologist gist plus CNNs (ligher blue or larger AUC).}
\label{fig:auc16}
\end{figure*}

Figures~\ref{fig:BothWrong}, ~\ref{fig:modelIntroducesError}, and ~\ref{fig:modelCorrectsError} show the imaging for which both humans and CNNs make mistakes, model introduced errors, and model corrected human errors. 
Figure~\ref{fig:auc16} shows the receiver operating characteristic curve (ROC curve) for each of the 16 end-to-end models between the CNNs and radiologist gist plus CNNs. Tables~\ref{tab:16accuracy} and ~\ref{tab:differences} show the AUC data and differences used in the main text from the 16 conditions we have tested.
Although the present work provides knowledge about adding radiologists' data into CNN, it does not allow for training on new data, thus restricting our application domain to a small range of in-house data.

Breast cancer is the second leading cause of cancer deaths in women in the developed countries with 11,000 breast cancer deaths in the UK and more than 40K in the US, annually, representing around $7\%$ and $12\%$ of all cancer deaths, our results may indicate that the global gist signal can substantially contribute to screening accuracy. 

\bibliography{HumanPlusAIIsBetter}

\section*{Author contributions statement}


A.S., J.M.W, J.C., and K.E conceived the study. S.W. wrote the software and performed the analysis. All assessed the results and worked on the manuscript.

\section*{Competing Interests}

The authors declare no competing interests. 


\end{multicols}

\begin{table}[ht]
\centering

\begin{adjustbox}{width=1\textwidth}
\small
\begin{tabular}{lllllll}
\hline
 & Accuracy & TPR & FPR & AUC & AUC stdev & 95\% CI \\
 \hline
Radiologist & 0.807 & 0.690 & 0.073 & 0.809 &  &  \\
InceptionV4 pretrained &  &  &  &  &  &  \\
LGBM &  &  &  &  &  &  \\
Preprocessing 1 & 0.687 &0.714 &0.341 &	0.743 & 0.055	&[0.628, 0.843] \\
Preprocessing 1 with gist & 0.771	&0.786&	0.244&	0.826&	0.047&	[0.726 ,0.910] \\
Preprocessing 2 & 0.614&	0.667&	0.439&	0.681&	0.060&	[0.558 ,0.789]
 \\
Preprocessing 2 with gist & 0.855&	0.857&	0.146&	0.899&	0.036&	[0.823, 0.962]
 \\
Preprocessing 3 & 0.566	&0.690&	0.561&	0.625&	0.062&	[0.501, 0.739]
 \\
Preprocessing 3 with gist & 0.663&	0.714&	0.390&	0.702&	0.056&	[0.591, 0.804]
 \\
Preprocessing 4 & 0.627	&0.690&	0.439	&0.676&	0.059&	[0.555, 0.784]
 \\
Preprocessing 4 with gist & 0.735&	0.714&	0.244&	0.814&	0.045&	[0.723, 0.899]
 \\
SVM &  &  &  &  &  &  \\
Preprocessing 1 & 0.578&	0.643&	0.488&	0.670&	0.058&	[0.554, 0.781]
 \\
Preprocessing 1 with gist & 0.795&	0.786&	0.195&	0.832&	0.048&	[0.735, 0.919]
 \\
Preprocessing 2 & 0.651&	0.643&	0.341&	0.639&	0.061&	[0.513, 0.755]
 \\
Preprocessing 2 with gist & 0.795&	0.786&	0.195&	0.839&	0.046&	[0.747, 0.925]
 \\
Preprocessing 3 & 0.602&	0.643&	0.439&	0.666&	0.060&	[0.542, 0.780]
 \\
Preprocessing 3 with gist & 0.807&	0.810&	0.195&	0.839&	0.046&	[0.739, 0.924]
 \\
Preprocessing 4 & 0.614&	0.667&	0.439&	0.666&	0.061&	[0.535, 0.774]
 \\
Preprocessing 4 with gist & 0.783&	0.762&	0.195&	0.843&	0.045&	[0.753, 0.926]
 \\
 &  &  &  &  &  &  \\
VGG19 pretrained &  &  &  &  &  &  \\
LBGM &  &  &  &  &  &  \\
Preprocessing 1 & 0.530&	0.548&	0.488	&0.554&	0.063&	[0.427, 0.674]
 \\
Preprocessing 1 with gist & 0.651&	0.738&	0.439&	0.761&	0.051&	[0.658, 0.851]
 \\
Preprocessing 2 & 0.639	&0.762&	0.488&	0.659&	0.061&	[0.525, 0.773]
 \\
Preprocessing 2 with gist & 0.711&	0.738&	0.317&	0.790&	0.047&	[0.701, 0.880]
 \\
Preprocessing 3 & 0.530&	0.595&	0.537&	0.516&	0.063&	[0.388, 0.640]
 \\
Preprocessing 3 with gist & 0.687&	0.714&	0.341&	0.761&	0.054&	[0.655, 0.861]
 \\
Preprocessing 4 & 0.627&	0.619&	0.366&	0.646&	0.060&	[0.526, 0.761]
 \\
Preprocessing 4 with gist & 0.723&	0.738&	0.293&	0.780&	0.050&	[0.678, 0.876]
 \\
SVM &  &  &  &  &  &  \\
Preprocessing 1 & 0.651	&0.667&	0.366&	0.701&	0.058&	[0.580, 0.814]
 \\
Preprocessing 1 with gist & 0.783&	0.762	&0.195&	0.850&	0.042&	[0.756, 0.925]
 \\
Preprocessing 2 & 0.723	&0.714&	0.268&	0.828&	0.044&	[0.739, 0.908]
 \\
Preprocessing 2 with gist & 0.819&	0.738&	0.098&	0.861&	0.041&	[0.776, 0.932]
 \\
Preprocessing 3 & 0.530&	0.571&	0.512&	0.553&	0.065&	[0.427, 0.681]
 \\
Preprocessing 3 with gist & 0.759&	0.762&	0.244&	0.807&	0.048&	[0.702, 0.891]
 \\
Preprocessing 4 & 0.651	&0.643&	0.341&	0.680&	0.056&	[0.570, 0.789]
 \\
Preprocessing 4 with gist & 0.783&	0.690&	0.122&	0.814&	0.047&	[0.717, 0.898]

\end{tabular}
\end{adjustbox}
\caption{The 16 end-to-end classificaiton methods with and without gist. The four preprocessing methods are as follows: (1) No changes, (2) Left breast images are horizontally mirrored so all images face the same direction, (3) Muscle fibers are cropped out of the original images, (4) Muscle fibers are cropped, and then left breast images are mirrored.}
\label{tab:16accuracy}
\end{table}

\begin{table}[ht]
\centering
\begin{adjustbox}{width=1\textwidth}
\small
\begin{tabular}{lllll}
 & Improvement TPR & Improvement FPR & Improvement AUC & Improvement Accuracy \\
  & (positive is good) &(negative is good) & (positive is good) & (positive is good) \\
InceptionV4+LGBM+preprocessing1 & 0.072&	-0.097&	0.083&	0.084
 \\
InceptionV4+LGBM+preprocessing2 & 0.190&	-0.293&	0.218&	0.241
 \\
InceptionV4+LGBM+preprocessing3 & 0.024&	-0.171&	0.077&	0.097
 \\
InceptionV4+LGBM+preprocessing4 & 0.024&	-0.195&	0.138&	0.108
 \\
InceptionV4+SVM+preprocessing1 & 0.143&	-0.293&	0.162&	0.217
 \\
InceptionV4+SVM+preprocessing2 & 0.143&	-0.146&	0.200&	0.144
 \\
InceptionV4+SVM+preprocessing3 & 0.167&	-0.244&	0.173&	0.205
 \\
InceptionV4+SVM+preprocessing4 & 0.095&	-0.244&	0.177&	0.169
 \\
VGG19+LGBM+preprocessing1 & 0.190&	-0.049&	0.207&	0.121
 \\
VGG19+LGBM+preprocessing2 & -0.024&	-0.171&	0.131&	0.072
 \\
VGG19+LGBM+preprocessing3 & 0.119&	-0.196&	0.245&	0.157
 \\
VGG19+LGBM+preprocessing4 & 0.119&	-0.073&	0.134&	0.096
 \\
VGG19+SVM+preprocessing1 & 0.095&	-0.171&	0.149&	0.132
 \\
VGG19+SVM+preprocessing2 & 0.024&	-0.170&	0.033&	0.096
 \\
VGG19+SVM+preprocessing3 & 0.191&	-0.268&	0.254&	0.229
 \\
VGG19+SVM+preprocessing4 & 0.047&	-0.219&	0.134&	0.132

\end{tabular}
\end{adjustbox}
\caption{Table showing the changes made to each of 16 model's results when incorporating radiologist response into the input vector. The value shows the difference $Result_{with\; gist} - Result_{without}$.}
\label{tab:differences}
\end{table}

\end{document}